# Image Inpainting by Adaptive Fusion of Variable Spline Interpolations

Zahra Nabizadeh, Ghazale Ghorbanzade, Nader Karimi, Shadrokh Samavi
*Isfahan University of Technology*
*Isfahan,* 84156-83111 *Iran,*

*Abstract*— There are many methods for image enhancement. Image inpainting is one of them which could be used in reconstruction and restoration of scratch images or editing images by adding or removing objects. According to its application, different algorithmic and learning methods are proposed. In this paper, the focus is on applications, which enhance the old and historical scratched images. For this purpose, we proposed an adaptive spline interpolation. In this method, a different number of neighbors in four directions are considered for each pixel in the lost block. In the previous methods, predicting the lost pixels that are on edges is the problem. To address this problem, we consider horizontal and vertical edge information. If the pixel is located on an edge, then we use the predicted value in that direction. In other situations, irrelevant predicted values are omitted, and the average of rest values is used as the value of the missing pixel. The method evaluates by PSNR and SSIM metrics on the Kodak dataset. The results show improvement in PSNR and SSIM compared to similar procedures. Also, the run time of the proposed method outperforms others.

## I. Introduction

Image inpainting is a computer vision technique that has a broad definition. It is a process that restores damaged or lost part of an image, insert an object into or remove an object from an image as the human eye could not understand it [1][2]. According to its definition, there are many applications for it; for example, it could be used for the restoration of old images, editing or composition of an image, etc. [1][3].

Most articles in this field can be classified into one of the two categories:
- Algorithmic approaches
- Learning approaches

### A. Algorithmic Approaches

Different techniques of image processing are used for image inpainting. In the following, the recent and essential procedures would be described.

Predicting missing pixels in an image are very depended on neighboring pixels. Accordingly, M.Akbari et al. proposed a method for selecting the best patch around a pixel, which leads to better prediction. By choosing this, the computational time reduced, and the output result is reasonable [4].

Interpolation is one of the image processing techniques that could predict pixels between pixels. In [5], M. Motmaen et al. utilize one and two-dimensional bicubic spline interpolation in multiple directions to predict missing pixels.

One of the applications of image inpainting is a combination of two images. In some situations, the source image has an inconsistent color, structure, and texture. Therefore, the output of the image inpainting has visible artifacts. In [6] S. Darabi et al. proposed a method that gradually transmits the features of the source image to the other image. They named this process image melding. The base of this process is patch-based optimization with three changes in it, (i) different geometric and photometric transformations are used to improve patch search space, (ii) a screened Poisson equation solver is used instead of the usual color averaging and (iii) using of new energy for reducing artifacts.

As it was mentioned before, selecting pixels, which are related to the missing pixel, is very effective in image inpainting. Exemplar-based algorithms are image inpainting techniques that have taken this information into consideration. In these algorithms, there are two main steps, (i) specifying the order of filling the lost block and (ii) finding good exemplars. In the previous works, the problem of selecting improper exemplar existed. L.Deng et al. [7] by investigating the propagation of patches, proposed a method in which the geometry and texture features have different priorities for transmitting to the output image.

In some techniques of image inpainting, finding matched patches, and in other mappings, similar patches to one bin are used. In [8] a method is proposed based on these two techniques and is named Coherency Sensitive Hashing.

There are two classes of algorithms for filling missed blocks in an image: (i) texture synthesis and (ii) image inpainting. The first class is used for large regions of missed blocks, and the second one is used for small missed blocks. In [9] by combining these two classes of algorithms, it has benefited from both.

### B. Learning approaches

Deep Neural Networks (DNN) have recently received much attention because of their capabilities in different applications with large datasets. Image inpainting is one of the applications that benefit from these neural networks. In most recent articles in this field, different types of DNN with different views are implemented. The following proposed DNN in some of these articles would be explained.

Y. Shin et al. proposed a new architecture for DNN to reduce computational resources and network parameters, which is the disadvantage of Generative Adversarial Network (GAN). Their proposed network PEPSI: parallel extended-decoder path for semantic inpainting address that problem by using a single shared encoding network and a parallel decoding network that has two coarse and inpainting paths [1].

In damaged imaged with large masked areas, lack of surrounding information of central pixels will cause blurred reconstructed images. In this case, it has seen that progressive methods have better results with less blurriness. That's why in [10], a full-resolution residual network (FRNN) has designed as a progressive network. In the proposed architecture, N Blocks, One Dilation strategy is adopted to achieve better quality. Also, a step-wise loss function is applied to improve intermediate restorations performance.

In most of the learning approaches, missing regions have been filled either by copying image patches or generating semantically-coherent patches. In [11], both of visual and semantical viewpoints are considered concurrently. To achieve this purpose, a Pyramid-context ENcoder Network (PEN-Net) has been used. By using such encoder, high-level semantic feature maps would be transferred to low-level feature maps and in a pyramid format, both semantic and visual coherency would be guaranteed. In addition, a multi-scale decoder is used to complete the network, with deeply-supervised pyramid losses and an adversarial loss. The proposed method leads to more realistic results and better performance. Also, in [12], advantages of traditional approaches that have better visual results near the damaged region boundaries are used with generation new structures, especially for central pixels. This is done through a feed-forward fully convolutional neural network.

In [13], a model called Deep Fusion Network (DFNet) has been proposed to address nonharmonic region boundaries. DFNet is an U-Net architecture embedded with introduced fusion blocks which applied in a multi-scale fashion. Each of the fusion blocks follows a feature map layer and is an alpha composition map of the resized input image and this feature map. A new measurement Boundary Pixel Error (BPE) besides such architecture, provides well performance in both qualitative and quantitative aspects.

One of the challenges in image inpainting is filling damaged areas which consist of one part of the foreground and one part of the background. In this case, the information of either being foreground or background is an important factor in predicting the missing pixel. In [1], this problem has been addressed. To this end, in the proposed method a contour predictor has been learned which would be used as the guidance of image inpainting block. In the contour predictor, foreground contours are extracted in the first step, and then, they have been completed in the next step. The output of completed contours would be used as the side information of the masked image to predict the missing values. In [14] also, the same problem as [1] is addressed. The proposed method contains a two-stage adversarial model EdgeConnect that continuously performs edge generation and image completion stages. By using such architecture, the quantity and quality of the performance would outperform state-of-the-art techniques.

In this paper, the focus is on the old scratched image reconstruction, which could be used in a scanner device or the device with limited resources. Therefore, a simple, fast and reasonable approach should be proposed for this application. In the remainder of this paper, the proposed method is illuminated in section II, experimental results are expressed in section III, and eventually, a concise conclusion is presented in section IV.

## II. PROPOSED METHOD

In natural images, there is less high-frequency data. Therefore, it is assumed that the pixels which are more close to each other, have more correlations. Accordingly, in image inpainting, missed pixels could be predicted by using surrounding pixels. In this paper, an image inpainting method is proposed which uses adaptive neighbor selection for spline interpolation. The spline interpolation method is selected for two reasons, (i) its less computational cost compared with deep neural networks and (ii) consistent results for small corrupted areas. Due to these two advantages, the proposed method could be used in devices with limited computational resources such as scanners for old and historical scratched images. In old photos, most of the missing pixels are only in lines, not in patches. The block diagram of the proposed method is shown in Fig. 1. Our approach contains three main blocks of pixel selection, directional interpolation, and fusion. In the following, these blocks will be explained in detail.

### A. Pixel Selection

The image inpainting process needs two inputs. One is the damaged image, and the other one is the mask, which specifies the location of lost pixels. By using this mask, only the lost pixels is selected for the process of image inpainting. This selection reduces the computational costs of the next blocks.

### B. Directional Interpolation

This block interpolates each pixel value in four directions, horizontal, vertical, 45, and 135 degrees diagonal. In this block for each pixel which received from the previous block, two tasks are done, (i) locality analysis for finding the best neighbors in each direction and (ii) spline interpolation. In the following, these two parts would be explained.

#### 1) Locality Analysis

In image inpainting, the selection of neighbor pixels around the missing pixel, which are involved in the prediction process, is an essential step. In [5] a fixed neighborhood around the missing pixel, and in [4], the best patch was used. In our work, according to the location of the missing pixel in the damaged part, appropriate neighboring pixels are adaptively selected. In Fig. 2, this process is shown for three scenarios. In this figure the missing pixel, which has to be predicted, is shown with red colo. The blue pixels, in Fig. 2, are the missing pixels around the selected pixel. Pixels that are selected for the prediction process are green. If the missing pixels after the selected pixel are more than the missing pixels before it, then more neighbors are chosen from the previous pixels. If missing pixels before the selected pixel are more, the neighboring pixels are selected from those that follow the selected pixel. In another situation, the same number of pixels are chosen from each side. The reason for selecting these neighbors is using the best correlation among the selected pixels. As it is shown in Fig. 1, in our approach, this selection is separately done for four directions.

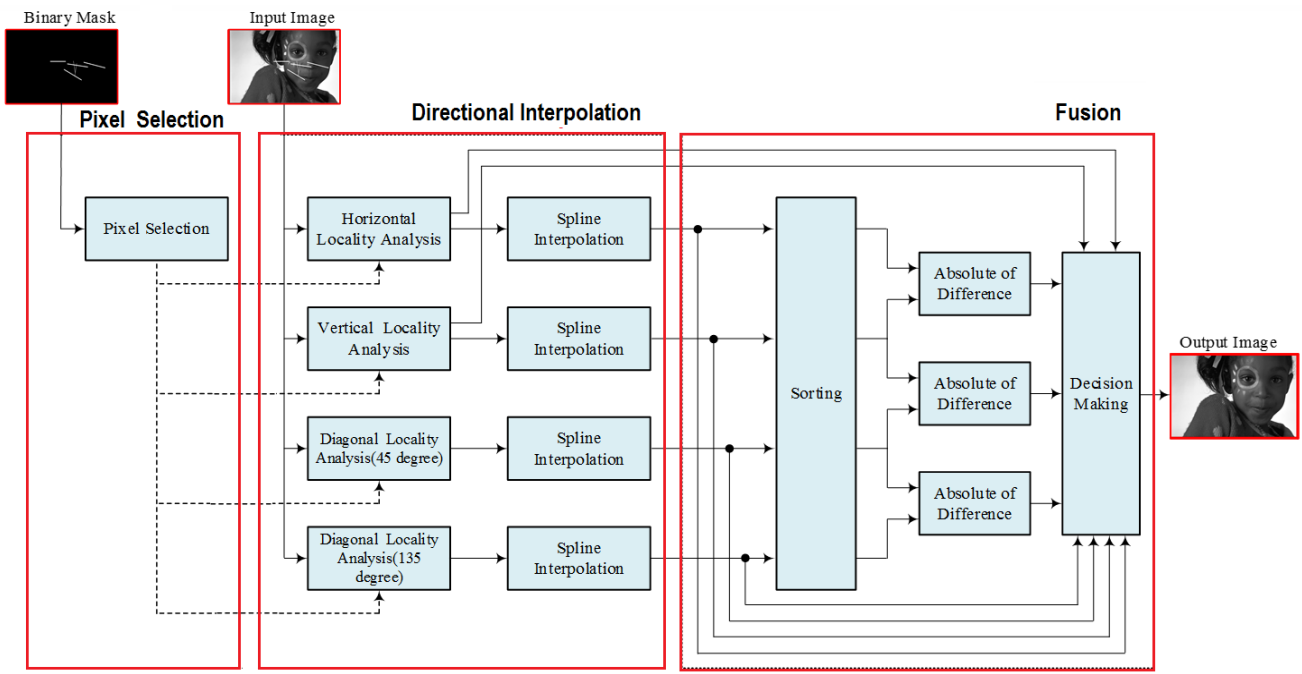

Figure 1: Block diagram of the proposed method

In addition to neighborhood selection, edge recognition is also important. In this paper, the horizontal and vertical edges are detected. In the horizontal direction, two auxiliary vectors with the same size of the missing pixels' neighbors are selected from top and bottom. By checking these three vectors we could detect that the lost pixel is on a horizontal edge. It is the same for vertical direction in which two vectors with the same size of lost pixels' neighbors are chosen from left and right. With this data, it could be detected that the lost pixel is on the vertical edge. If the pixels of auxiliary vectors are the members of the missing block, they are omitted. If this deletion is not performed, the wrong edge is detected. In Fig. 3, the selection of these auxiliary vectors is shown. The red pixel is lost pixel, the green pixels are lost pixel's neighbors, and the magenta pixels are auxiliary pixels for edge detection.

*2) Spline Interpolation*

Spline interpolation is a method for finding the intermediate value of data by fitting a curve on the existing data. In our work, this method is used to interpolate the lost pixels from their neighbors. There are two approaches for interpolating the lost pixels, (i) pixel-wise interpolation and (ii) interpolate the block of lost pixels. In the latter approach, because of fixed neighbor for all pixels in the lost block, the output results are not accurate enough. So, in this paper due to the adaptive neighbor for each pixel, the first approach is used. For each lost pixel, in this block, the spline interpolation predicts the value of the pixel by using the information from the previous block in four directions.

*C. Fusion*

By following the previous blocks, for each lost pixel, four predicted values are produced. The purpose of this block is finding the best combination of the four previous blocks' output for each pixel. This block consists of three steps, (i) sorting the four values for each pixel, (ii) calculate the difference of them, and (iii) produce proper output. The details of these three steps would be explained.

According to the direction of the missing blocks, the predicted result, which is in the same direction as the missing block is not valid. For finding the invalid result and removing it from the result set, in first step, by considering the information of edge detection, if the lost pixel is on the vertical edge, then the predicted value of vertical is used as valid result and the other results are invalid and if the lost pixel is on the horizontal edge, then the predicted value of horizontal is used as valid result and the other results are invalid, in other situation the pixel is the smooth area and the value of lost pixel select according to the following description. At first, the results are sorted. The output of this step is a sorted vector with four elements, which is named $'V'$. In the second step, the difference of $(V(1), V(2))$, $(V(2), S\_V(3))$, and $(V(3), V(4))$ are calculated. In the last step, by using the difference values, selecting the invalid result are three modes (i) If the diff($V(1)$, $V(2)$) is the maximum value among the difference values, V(1) is invalid result, (ii) If the diff(V(3), V(4)) is the maximum value among the difference values, V(4) is invalid result and (iii) in other situation there is no distinct reason for selecting invalid result, so all four results would be used. In the end, the average of selected valid values is used as the predicted value for the missing pixel.

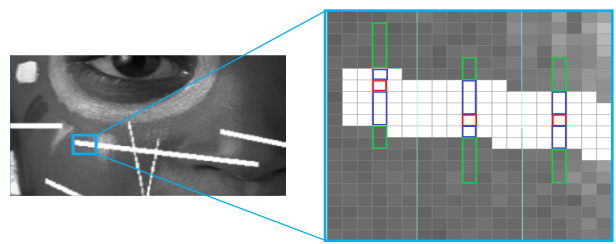

Figure 2: Three scenarios of selecting adaptive neighbors

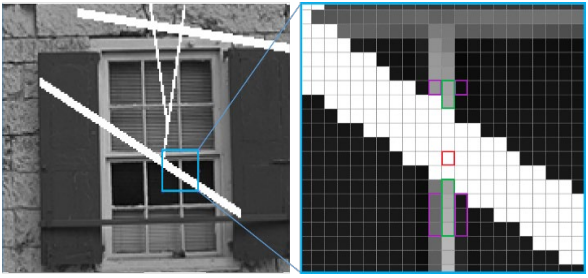

Figure 4: Auxiliary vectors for edge detection

## III. EXPERIMENTAL RESULTS

### A. Implementation Details

As mentioned before, mask is one of the inputs of image inpainting techniques. The locations of missing pixels could be a challenge for different image inpainting techniques. The missing pixels could be in a textured area or on an edge. For this purpose, in a zero matrix, thick lines with random lengths and locations are drawn. Then this mask is applied to images and the damaged images are created as another input of the image inpainting system. The code of this method is implemented with MATLAB R2016b and runs on the Core i5 CPU @2.5GHz with 4 GB RAM.

### B. Comparison with other Methods

The proposed method compares with three similar techniques, which are proposed in [4], [5] and [9]. To evaluate the results, all these methods test with 15 images, which are from Kodak dataset. Three metrics Peak Signal to Noise Ratio (PSNR), Structural Similarity (SSIM) and times are calculated for comparison. The average PSNR is given in Table 1. The results show that adaptive selecting of neighbors in four dimensions, detecting the location of pixel on edge, and smart fusion of the results improve the average of PSNR. The results of SSIM and times for our method and the methods of [4] and [5] are reported in Table 2. By preserving the structures of the images the run time of our method is very short compared to other methods. In [5], two-dimensional splines are used, but in our method, by considering adaptive neighbors, one-dimensional splines are sufficient. In [4] for finding the best patch the whole image is searched, but in our approach, only the neighbors of lost pixels are considered. To show the situation that our method predicts missing pixels better than the outputs of the method in [5], one of the output are displayed visually in . In Fig. 4, the damaged image, the output of our method, and the method implemented in [5] are shown. As the figure shows, by considering the location of the lost pixel on edge, the predicted value could be near to its real value.

## IV. CONCLUSION

In this paper, a new procedure for image inpainting is proposed. Our focus is on the application of removing scratches from old images. Such applications, most of the time, are implemented in devices with limited computational resources, such as scanners or cameras. Compared with the previous methods, this paper has offered three contributions: (i) adaptive selection of neighbors of the missing pixel, (ii) detection of the location of the missing pixel on an edge, and (iii) removing invalid results and using the average value of the valid results. Our results show improvements in all metrics, especially the execution time of the method. For better results, we could also consider edges in 45 and 135 degrees and implement an adaptive selection of thresholds for edge detection.

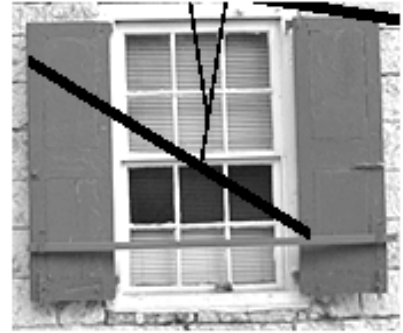

(a)

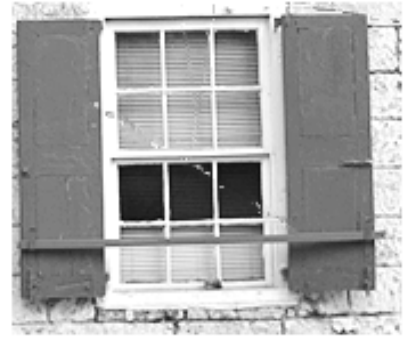

(b)

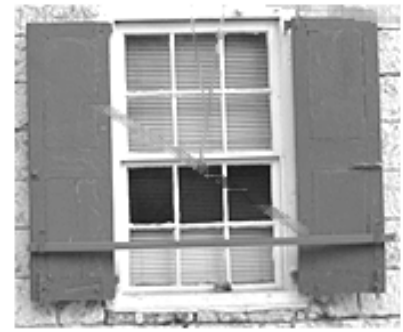

(c)

Figure 3: (a) The damaged image, (b) output of our method, and (c) output of [5]

Table 1: PSNR comparison of proposed methods with the proposed method in [9], [4] and [5]

| Metric | Criminisi [9] | RIBBONS [4] | Motmaen [5] | Ours |
|---|---|---|---|---|
| PSNR | 43.56 | 31.29 | 44.14 | 44.22 |

Table 2: The results for SSIM and times of our method and the methods in [4] and [5]

| Metric | RIBBONS [4] | Motmaen [5] | Ours |
|---|---|---|---|
| SSIM | 0.9776 | 0.9965 | 0.9966 |
| Times (sec) | 5.52 | 12.56 | 3.36 |